\documentclass[aps,prd,twocolumn,superscriptaddress,amssymb,eqsecnum,showpacs,showkeyes,secnumarabic,graphics,floatfix,nofootinbib,tightenlines,longbibliography]{revtex4-1}

\usepackage{graphicx}
\usepackage{epstopdf}
\usepackage{epsfig}
\usepackage{bm}
\usepackage{cancel}
\usepackage{dcolumn}
\usepackage{amsmath}
\usepackage{hhline}
\usepackage[utf8]{inputenc}
\usepackage{multirow}
\usepackage[breaklinks=true,colorlinks=true,
linkcolor=blue,urlcolor=blue,citecolor=cyan,
bookmarks=true,bookmarksopenlevel=2]{hyperref}

\def \beq  {\begin{equation}}
\def \eeq  {\end{equation}}
\def \ber  {\begin{eqnarray}}
\def \eer  {\end{eqnarray}}
\def \lleq {\lower0.9ex\hbox{ $\buildrel < \over \sim$} ~}
\def \ggeq {\lower0.9ex\hbox{ $\buildrel > \over \sim$} ~}

\def \beq  {\begin{equation}}
\def \eeq  {\end{equation}}
\def \ber  {\begin{eqnarray}}
\def \eer  {\end{eqnarray}}
\def\bq{\begin{equation}}
\def\nq{\end{equation}}
\def\bqr{\begin{eqnarray}}
\def\nqr{\end{eqnarray}}

\begin{document}
\newcommand{\newc}{\newcommand}

\newc{\be}{\begin{equation}}
\newc{\ee}{\end{equation}}
\newc{\ba}{\begin{eqnarray}}
\newc{\ea}{\end{eqnarray}}
\newc{\bea}{\begin{eqnarray*}}
\newc{\eea}{\end{eqnarray*}}
\newc{\D}{\partial}
\newc{\ie}{{\it i.e.} }
\newc{\eg}{{\it e.g.} }
\newc{\etc}{{\it etc.} }
\newc{\etal}{{\it et al.}}
\newc{\lcdm}{$\Lambda$CDM }
\newcommand{\nn}{\nonumber}
\newc{\ra}{\Rightarrow}

\title{Sudden Future Singularities in Quintessence and Scalar-Tensor Quintessence Models}
\author{A. Lymperis}\email{alymperis@upatras.gr} 
\affiliation{Department of Physics, University of Patras, 26500 Patras, Greece}
\author{L. Perivolaropoulos}\email{leandros@uoi.gr}
\affiliation{Department of Physics, University of Patras, 26500 Patras, Greece}
\affiliation{on leave from Department of Physics, University of Ioannina, Ioannina 45110, Greece}
\author{S. Lola}\email{ magda@physics.upatras.gr } 
\affiliation{Department of Physics, University of Patras, 26500 Patras, Greece}

\date {\today}

\begin{abstract}
We demonstrate analytically and numerically the existence of geodesically complete singularities in quintessence and scalar tensor quintessence models with scalar field potential of the form $V(\phi)\sim \vert \phi\vert^n$ with $0<n<1$. In the case of quintessence, the singularity which occurs at $\phi=0$, involves divergence of the third time derivative of the scale factor (Generalized Sudden Future Singularity (GSFS)), and of the second derivative of the scalar field. In the case of scalar-tensor quintessence with the same potential, the singularity is stronger and involves divergence of the second derivative of the scale factor (Sudden Future Singularity (SFS)). We show that the scale factor close to the singularity is of the form $a(t)=a_s+b(t_{s}-t) + c(t_{s}-t)^2 +d(t_{s}-t)^q$ where $a_s,b,c,d$ are constants obtained from the dynamical equations and $t_s$ is the time of the singularity. In the case of quintessence we find $q=n+2$ (\ie $2<q<3$), while for the case of scalar-tensor quintessence $q=n+1$ ($1<q<2$). We verify these analytical results numerically and  extend them to the case where a perfect fluid, with a constant equation of state $w=\frac{p}{\rho}$, is present. The linear and quadratic terms in $(t_{s}-t)$ are subdominant for the diverging derivatives close to the singularity, but can play an important role in the estimation of the Hubble parameter. Using the analytically derived relations between these terms, we derive relations involving the Hubble parameter close to the singularity, which may be used as observational signatures of such singularities in this class of models. For quintessence with matter fluid, we find that close to the singularity $\dot H=\frac{3}{2}\Omega_{0m} (1+z_{s})^{3}-3H^{2}$. These terms should be taken into account when searching for future or past time such singularities, in cosmological data.
\end{abstract}
\maketitle

\section{Introduction}
\label{sec:Introduction}

The fact that the Universe has entered a phase of accelerating expansion ($\ddot a>0$) \cite{Copeland:2006wr, Frieman:2008sn} has created new possibilities in the context of the study of exotic physics on cosmological scales. Cosmological observations of Type Ia supernova \cite{Riess:1998cb}, which were later supported by the cosmic microwave backround (CMB) \cite{Jaffe:2000tx} and the large scale structure observations \cite{Hawking:1973uf, Davis:1985rj}, are consistent with the existence of a cosmological constant (\lcdm model) \cite{Bull:2015stt} as the possible cause of this mysterious phenomenon. Despite the simplicity of \lcdm and its consistency with most cosmological observations \cite{Riess:1998cb} the required value of the cosmological constant needs to be fine-tuned in comparison with microphysical expectations. This problem has lead to the consideration of models alternative to \lcdm. Such models include modifications of GR \cite{Nojiri:2006ri, Nojiri:2003ft}, scalar field dark energy (quintessence) \cite{Zlatev:1998tr, Carroll:1998zi}, physically motivated forms of fluids \eg Chaplygin gas \cite{Bento:2002ps, Bilic:2001cg} etc. \par
Some of these dark energy models predict the existence of exotic cosmological singularities, involving divergences of the scalar spacetime curvature and/or its derivatives. These singularities can be either  geodesically complete \cite{Scherrer:2004eq, Nesseris:2004uj, Perivolaropoulos:2004yr, Lykkas:2015kls} (geodesics continue beyond the singularity and the Universe may remain in existence) or geodesically incomplete \cite{Dabrowski:2014fha} (geodesics do not continue beyond the singularity and the Universe ends at the classical level).
They appear in various physical theories such as superstrings \cite{Antoniadis:1993jc}, scalar field quintessence with negative potentials\cite{Felder:2002jk}, modified gravities and others \cite{Lykkas:2015kls, Barrow:2004xh, FernandezJambrina:2004yy}. Violation of the cosmological principle (isotropy-homogeneity) by some cosmological models (\eg modified gravity \cite{Mota:2006fz}, quantum effects \cite{Onemli:2004mb}), has been shown to eliminate or weaken both geodesically complete and incomplete singularities \cite{Fewster:2010gm, Zhang:2011uv, Bamba:2012ky, Nojiri:2010pw, Bamba:2012vg, Barrow:2011ub, BouhmadiLopez:2009jk, Bouhmadi-Lopez:2014jfa, BouhmadiLopez:2009pu, Kamenshchik:2015oya, Kamenshchik:2013gza, Kamenshchik:2012ij, Dabrowski:2006dd, Dabrowski:2013sea, FernandezJambrina:2008dt, Kamenshchik:2007zj, Nojiri:2008fk, Nojiri:2009pf, Sami:2006wj, Singh:2010qa}. \par
Geodesically incomplete singularities include the Big-Bang \cite{Penrose:1988ph}, the Big-Rip \cite{Briscese:2006xu, Chimento:2004ps} where the scale factor diverges at a finite time due to infinite repulsive forces of phantom dark energy, the Little-Rip \cite{Frampton:2011rh} and the Pseudo-Rip \cite{Frampton:2011aa} singularities where the scale factor diverges at a infinite time and the Big-Crunch \cite{Felder:2002jk, Heard:2002dr, Elitzur:2002rt, Giambo:2015tja, Perivolaropoulos:2004yr, Lykkas:2015kls} where the scale factor vanishes due to the strong attractive gravity of future envolved dark energy, as \eg in quintessence models with negative potential. \par
Geodesically complete singularities include SFS (Sudden Future Singularity) \cite{Barrow:2004xh}, FSF (Finite Scale Factor) \cite{bouhmadi2008worse}, BS (Big-Separation) and the w-singularity \cite{FernandezJambrina:2010ck}. In these singularities, the cosmic scale factor remains finite but a scale factor's derivative diverges at a finite time. The singular nature of these ``singularities" amounts to the divergence of scalar quantities involving the Riemann tensor and the Ricci scalar $R=6\left (\frac{\ddot {a}}{a}+\frac{\dot a^{2}}{a^{2}}+\frac{k}{a^{2}} \right )$, for the FRW metric, where $a(t)$ is the cosmic scale factor \cite{FernandezJambrina:2006hj}. Despite the divergence of the Ricci scalar, the geodesics are well defined at the time of the singularity. The Tipler and Krolak \cite{Tipler:1977zza, Krolac1986} integrals of the Riemann tensor components along the geodesics are indicators of the strength of these singularities and remain finite in most cases. The Tipler integral \cite{Tipler:1977zza} is defined as

\be
\int^{\tau}_{0}d\tau'\int^{\tau}_{0}d\tau'|R^{i}_{0j0}(\tau'')|
\ee

while the Krolak integral \cite{Krolac1986} is defined as

\be
\int^{\tau}_{0}d\tau'|R^{i}_{0j0}(\tau'')|
\ee

\noindent where $\tau$ is the affine parameter along the geodesic and $R^{i}_{0j0}$ is the Riemann tensor. The components of the Riemann tensor are expressed in a frame that is parallel transported along the geodesics. If the scale factor's first derivative is finite at the singularity, both integrals are finite (even if the second derivative of the scale factor diverges), since the Riemann tensor components involve up to second order derivatives of the scale factor. If, however, the first derivative of the finite scale factor diverges, then it is easy to see from the above integrals that only the Tipler integral is finite at the singularity, while the Krolak integral diverges. This implies an infinite impulse on the geodesics, which dissociate all bound systems at the time of the singularity \cite{Perivolaropoulos:2016nhp,Nesseris:2004uj}. The singularities that lead to the divergence of the above integrals are defined as {\it strong singularities} \cite{Rudnicki:2002ep, Rudnicki:2006hu}. \par
It is interesting to connect these singularities with the properties of the cosmic energy-momentum tensor. In FRW spacetime with metric

\be
ds^2=-dt^2+a^{2}(t)\bigg[\frac{dr^2}{1-kr^2}+r^2(d\theta^2+\sin^{2}\theta d\phi^2)\bigg]
\ee

\noindent we assume standard Einstein-Hilbert action

\be \label{action}
\mathcal{S}=\int \left [\frac{1}{16\pi G}R+\mathcal{L}_{(fluid)} \right ]\sqrt{-g}d^{4}x
\ee

The Friedmann equations obtained by variation of the above action connect the density and pressure with the cosmic scale factor $a(t)$:

\be \label{density}
\rho(t)=\frac{3}{8\pi G} \left (\frac{\dot{a}^{2}}{a^{2}}+\frac{k}{a^{2}} \right )
\ee

\be \label{pressure}
p(t)=-\frac{1}{8\pi G} \left (2\frac{\ddot{a}}{a}+\frac{\dot{a}^{2}}{a^{2}}+\frac{k}{a^{2}} \right )
\ee

\noindent where the density and pressure are connected by the continuity equation:

\be \label{cont}
\dot \rho(t)=-3\frac{\dot{a}}{a} \left [\rho(t)+p(t) \right ].
\ee

\noindent In what follows we set $8\pi G=c=1$ and assume spatial flatness ($k = 0$), in agreement with observational results and WMAP \cite{Hinshaw:2012aka}.

The divergence of the scale factor and/or its derivatives leads to divergence of scalar quantities like the Ricci scalar thus to different types of singularities or `cosmological milestones' \cite{FernandezJambrina:2006hj}. However geodesics do not necessarily end at these singularities and if the scale factor remains finite they are extended beyond these events \cite{FernandezJambrina:2004yy} even though a diverging impulse may lead to dissociation of all bound systems in the Universe at the time $t_s$ of these events\cite{Perivolaropoulos:2016nhp}.  

Thus singularies can be classified \cite{PhysRevD.71.063004} according to the behaviour of the scale factor $a(t)$, and/or its derivatives at the time $t_s$ of the event or equivalently (according to eqs (\ref{density}),(\ref{pressure})) and the energy density and pressure of the content of the universe at the time $t_s$. A classification of such singularities and their properties is shown in Table \ref{TabI}.

A particularly interesting type of singularity is the Sudden Future Singularity \cite{Barrow:2004xh} which involves violation of the dominant energy condition
and divergence of the cosmic pressure, of the Ricci Scalar and of the second time derivative of the cosmic scale factor. The scale factor can be parametrized as 

\be \label{scalefactor0}
a(t)=\left (\frac{t}{t_{s}} \right )^{m} (a_{s}-1)+1-\left (1-\frac{t}{t_{s}} \right )^{q},
\ee

\noindent where $m, q, t_{s}$ are constants to be determined, $a_{s}$ is the scale factor at the time $t_{s}$ and $1<q<2$. For this range of the parameter $q$, according to eq. (\ref{density}), $a, \dot a$ and $\rho$ remain finite at $t_{s}$.
However, from eqs (\ref{pressure}), (\ref{cont}) it follows that $p, \dot \rho$ and $\ddot a$ become infinite. Thus, when the first derivative of the scale factor is finite at the singularity, but the second derivative diverges (SFS singularity \cite{Barrow:2004xh}), the energy density is finite but the pressure diverges. \par

Geodesically complete singularities where the scale factor behaves like eq. (\ref{scalefactor0}), are obtained in various physical models such as, anti-Chaplygin gas \cite{Kamenshchik:2013ink, Keresztes:2012zn}, loop quantum gravity \cite{Singh:2010qa}, tachyonic models \cite{Gorini:2003wa, Kamenshchik:2012ij, Kamenshchik:2013gza, Kamenshchik:2015oya}, brane models \cite{1126-6708-2003-10-066, Calcagni:2004bh, BouhmadiLopez:2009jk} etc. Such singularities however have not been studied in detail in the context of the simplest dark energy models of quintessence and scalar-tensor quintessence (see however \cite{Barrow:2015sga} for a qualitative analysis in the case of quintessence).

In Ref. \cite{Barrow:2015sga} it was shown through a qualitative analysis that a singularity of the GSFS type (see Table \ref{TabI}), involving a divergence of the third derivative of the scale factor, occurs generically in quintessence models with potential of the form

\be \label{potential}
V(\phi)=A|\phi|^{n},\ \ \ \ \  A>0,
\ee

\noindent with $0<n<1$ and $A$ a constant parameter. This is in fact the simplest extension of \lcdm with geodesically complete cosmic singularities and occurs at the time $t_s$ when the scalar field becomes zero ($\phi=0$). 

\begin{widetext}
\centering

\begin{table}
\caption{Classification and properties of cosmological singularities. The singularities discussed in the present analysis are indicated in bold.}\label{TabI}
\begin{tabular}{c c c c c c c c c c} \\
  \hline
 Name & $t_{sing}$ & $a(t_{s})$ & $\rho(t_{s})$ & $p(t_{s})$ & $\dot p(t_{s})$ & $w(t_{s})$ & T & K & Geodesically \\ 
 \hline\hline
 Big-Bang (BB) & 0 & 0 & $\infty$ & $\infty$ & $\infty$ & finite & strong & strong &  incomplete \\ 
 \hline
 Big-Rip (BR) & $t_{s}$ & $\infty$ & $\infty$ & $\infty$ & $\infty$ & finite & strong & strong & incomplete \\
 \hline
 Big-Crunch (BC) & $t_{s}$ & 0 & $\infty$  & $\infty$ & $\infty$ & finite & strong & strong &  incomplete \\
 \hline
 Little-Rip (LR) & $\infty$ & $\infty$ & $\infty$ & $\infty$ & $\infty$ & finite & strong & strong & incomplete \\
 \hline
 Pseudo-Rip (PR) & $\infty$ & $\infty$ & finite & finite & finite & finite & weak & weak & incomplete \\
 \hline
 \bf Sudden Future (SFS) & $\bf t_{s}$ & $\bf a_{s}$ & $\bf \rho_{s}$ & $\bf \infty$ & $\bf \infty$ &\bf finite &\bf weak & \bf weak &\bf complete \\  
 \hline
 Finite Sudden Future (FSF) & $t_{s}$ & $a_{s}$ & $\infty$ & $\infty$ & $\infty$ & finite & weak & strong & complete \\
 \hline
\bf Generalized Sudden Future (GSFS) & $\bf t_{s}$ & $\bf a_{s}$ & $\bf \rho_{s}$ & $\bf p_{s}$ & $\bf \infty$ &\bf finite & \bf weak & \bf strong & \bf complete \\
 \hline
 Big-Separation (BS) & $t_{s}$ & $a_{s}$ & 0 & 0 & $\infty$ & $\infty$ & weak & weak & complete \\
 \hline
 w-singularity (w) & $t_{s}$ & $a_{s}$ & 0 & 0 & 0 & $\infty$ & weak & weak & complete \\
 \hline
\end{tabular}
\end{table}
\end{widetext}

In the present study we extend the analysis of \cite{Barrow:2015sga} in the following directions:

\begin{enumerate}

\item{
We verify the existence of the GSFS both numerically and analytically, using a proper generalized expansion ansatz for the scale factor and the scalar field close to the singularity. This generalized ansatz includes linear and quadratic terms, that dominate close to the singularity and cannot be ignored when estimating the Hubble parameter and the scalar field energy density. Thus, they are important when deriving the observational signatures of such singularities.}
\item{
We derive analytical expressions for the power (strength) of the singularity in terms of the power $n$ of the scalar field potential.}
\item{
We extend the analysis to the case of scalar tensor quintessence with the same scalar field potential and derive both analytically and numerically the power of the singularity in terms of the power $n$ of the scalar field potential.}

\end{enumerate}  

The structure of this paper is the following: In section II we focus on the quintessence model of eq. (\ref{potential}), and investigate the strength of the GSFS both analytically and numerically. In section III we extend the analysis to the case of scalar tensor quintessence and investigate the modification of the strength of the singularity both analytically (using a proper expansion ansatz) and numerically, by explicitly solving the dynamical cosmological equations. Finally, in section IV we summarise our results and discuss possible extensions of the present analysis.  

\section{Sudden Future Singularities in Quintessence Models}
\label{sec:Section 2}

\subsection{Evolution without perfect fluid}

Setting $8\pi G=1$, the most general action, involving gravity, nonminimally coupled with a scalar field $\phi$, and a perfect fluid is

\be \label{action1}
\resizebox{0.49\textwidth}{!}
{
$\mathcal{S}=\int \left [\frac{1}{2}F(\phi)R+\frac{1}{2}g^{\mu \nu} \phi_{;\mu}  \phi_{;\nu}-V(\phi)+\mathcal{L}_{(fluid)} \right ]\sqrt{-g}d^{4}x.$
}
\ee

In the special case where $F(\phi)=1$ and in the absence of a perfect fluid, the action (\ref{action1}) reduces to the simple case of quintessece models without a perfect fluid

\be \label{action2}
\mathcal{S}=\int \left [\frac{1}{2}R+\frac{1}{2}g^{\mu \nu} \phi_{;\mu}  \phi_{;\nu}-V(\phi) \right ]\sqrt{-g}d^{4}x
\ee

The energy density and pressure of the scalar field $\phi$, may be written as 

\begin{center}
$\rho_{\phi}=\frac{1}{2}\dot \phi ^{2}+V(\phi)$ \ \ and \ \ $p_{\phi}=\frac{1}{2}\dot \phi ^{2}-V(\phi)$.
\end{center}

\noindent Variation of the action (\ref{action2}) assuming a power law potential (\ref{potential}) leads to the dynamical equations

\be \label{barrow1}
3H^2=\frac{1}{2} \dot{\phi}^{2} +V(\phi)
\ee

\be \label{barrow2}
\ddot \phi=-3H\dot{\phi}-An|\phi|^{n-1} \Theta(\phi)
\ee

\be \label{barrow3}
2\dot H=-\dot{\phi}^{2},
\ee

\noindent where $H=\frac{\dot a}{a}$ is the Hubble parameter, $0<n<1$ and 
\be 
\Theta(\phi)=
\begin{cases}
1, & \phi>0 \\
-1, & \phi<0
\end{cases}
\ee. 
This class of quintessence models has been studied extensively focusing mostly on the cosmological effects and the dark energy properties that emerge due to the expected oscillations of the scalar field around the minimum of its potential \cite{Perivolaropoulos:2002pn, Perivolaropoulos:2003we, Johnson:2008se-osc-pot, Lima:2013rja-osc-pot, Dutta:2008px-osc-pot}. In the present analysis we focus instead on the properties of the cosmological singularity that is induced as the scalar field vanishes periodically during its oscillations. For simplicity, we consider only the first time $t_s$ when the scalar field vanishes during its dynamical oscillations.

The dynamical evolution of the scalar field due to the potential shown in Fig. \ref{fig:fig1} may be qualitatively described as follows \cite{Barrow:2015sga}:

\begin{figure}[!h]
\centering
\vspace{0.3cm}\rotatebox{0}{\vspace{0cm}\hspace{0cm}\resizebox{0.48\textwidth}{!}{\includegraphics{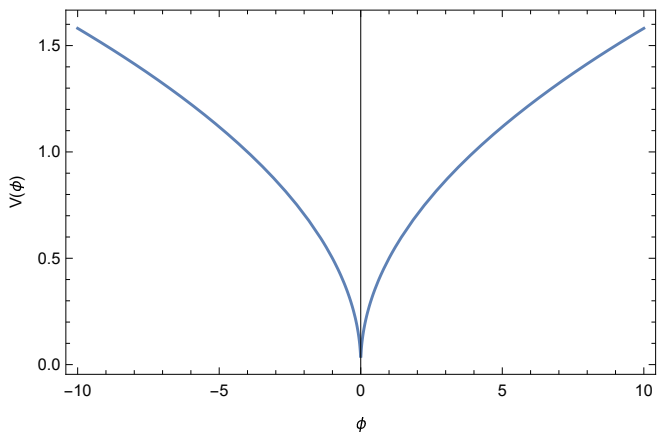}}}
\caption{Exponential scalar field potential $V(\phi)=A|\phi|^{n}$}
\label{fig:fig1}
\end{figure}

\noindent  From eqs (\ref{barrow1}), (\ref{barrow3}), it follows that when $t\to t_{s}$ ($\phi \to 0$) $H, \dot H$ remain finite and so does $\dot \phi$. But in eq. (\ref{barrow2}) there is a divergence of the term $\phi^{n-1}$ for $0<n<1$ and thus $\ddot \phi \to \infty$ as $\phi \to 0$. $\ddot H$ also diverges at this point due to the divergence of $\ddot \phi$, as follows by differentiating eq. (\ref{barrow3}). This implies that the third derivative of the scale factor diverges, and a GSFS occurs at this point (\ie $a_{s}, \rho_{s}, p_{s}$ remain finite but $\dot p \to \infty)$. Thus, the constraints on the power exponents $q,r$ of the diverging terms in the expansion of the scale factor ($\sim (t_{s}-t)^q$ ) and of the scalar field ($\sim (t_{s}-t)^r$ ) are $2<q<3$ and $1<r<2$ respectively (see eqs (\ref{scalefactor1}), (\ref{scalarfield}) below). 

In what follows we extend the above qualitative analysis to a quantitative level. In particular, we use a new ansatz for the scale factor and the scalar field, containing linear and quadratic terms of $(t_{s}-t)$. These terms play an important role since they dominate in the first and second derivative of the scale factor as the singularity is approached.  

Thus, the new ansatz for the scale factor which generalizes (\ref{scalefactor0}), by introducing linear and quadratic terms in $(t_{s}-t)$ is of the form

\be \label{scalefactor1}
a(t)=a_{s}+b(t_{s}-t)+c(t_{s}-t)^{2}+d(t_{s}-t)^{q},
\ee

\noindent where $b, c, d$ are real constants to be determined, and $2<q<3$ so that $\dddot a$ diverges at the GSFS.

\noindent The corresponding expansion of the scalar field $\phi(t)$ close to singularity is of the form

\be \label{scalarfield}
\phi(t)=f(t_{s}-t)+h(t_{s}-t)^{r}
\ee

\noindent where $f,h,$ are real constants to be determined, and $1<r<2$ so that $\ddot \phi$ diverges at the singularity. 

Substituting eqs (\ref{scalefactor1}), (\ref{scalarfield}) in eq. (\ref{barrow2}), we get the equation of the dominant terms

\be \label{intermediate}
\mathcal{A}_{1} (t_{s}-t)^{r-2}= \mathcal{A}_{2} (t_{s}-t)^{n-1}
\ee

\noindent where the $\mathcal{A}_{1}, \mathcal{A}_{2}$, denote constants, which may be expressed in terms of $f, h$ and the constant $A$ (see Appendix).
\noindent Clearly, both the left and right-hand side of eq. (\ref{intermediate}) diverge at the singularity for $1<r<2$ and $0<n<1$. Equating the power laws of divergent terms we obtain
\be \label{r}
r=n+1
\ee
Similarly, differentiation of eq. (\ref{barrow3}) with respect to $t$ gives $2\ddot H=-2\dot{\phi}\ddot{\phi}$, from which we obtain an equation for
the dominant terms using eqs (\ref{scalefactor1}), (\ref{scalarfield})

\be \label{intermediate1}
\mathcal{A'}_{1}(t_{s}-t)^{q-3}=\mathcal{A'}_{1}(t_{s}-t)^{r-2}
\ee

\noindent where the $\mathcal{A'}_{1}, \mathcal{A'}_{2}$ are constants, which may be expressed in terms of $d, f, h$ (see Appendix).
\noindent The left and the right-hand side of eq. (\ref{intermediate1}) diverge, 
and therefore, equating the power laws of diverging terms we obtain

\be \label{q}
q=r+1.
\ee

\noindent Thus, using (\ref{r}) and (\ref{q}) we find the exponent $q$ in terms of $n$ as

\be \label{q1}
q=n+2.
\ee

\begin{figure}[!h]
\centering
\vspace{0.3cm}\rotatebox{0}{\vspace{0cm}\hspace{0cm}\resizebox{0.49\textwidth}{!}{\includegraphics{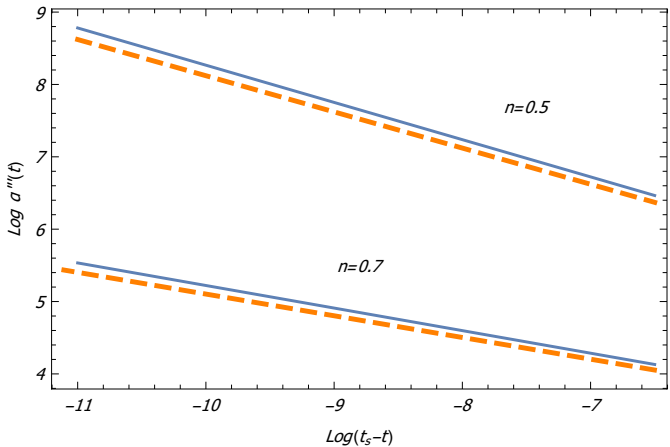}}}
\caption{Numerical verification of the $q$-exponent for $n=0.5$ and $n=0.7$. The orange dashed line, denotes the analytical, while the blue line denotes the numerical solution. As expected the slopes for each n are identical.}
\label{fig:fig2}
\end{figure}

\begin{figure}[!h]
\centering
\vspace{0.3cm}\rotatebox{0}{\vspace{0cm}\hspace{0cm}\resizebox{0.49\textwidth}{!}{\includegraphics{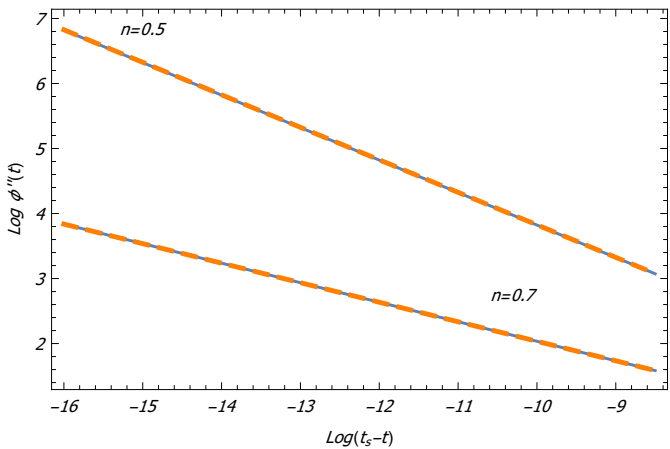}}}
\caption{Same as Fig. \ref{fig:fig2} for the $r$-exponent.}
\label{fig:fig3}
\end{figure}

Eqs (\ref{r}), (\ref{q1}), are consistent with the qualitatively expected range of $r,q$, for $0<n<1$.

\begin{figure}[!h]
\centering
\vspace{0.3cm}\rotatebox{0}{\vspace{0cm}\hspace{0cm}\resizebox{0.49\textwidth}{!}{\includegraphics{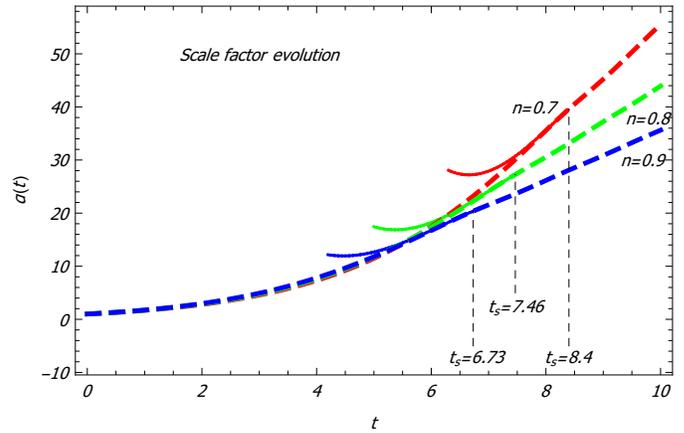}}}
\caption{Plot of numerical (dashed) and analytical (line) time evolution of the scale factor, for $n=0.7, 0.8, 0.9$. The two solutions for each $n$ are consistent close to each singularity.}
\label{fig4}
\end{figure}

\begin{figure}[!h]
\centering
\vspace{0.3cm}\rotatebox{0}{\vspace{0cm}\hspace{0cm}\resizebox{0.49\textwidth}{!}{\includegraphics{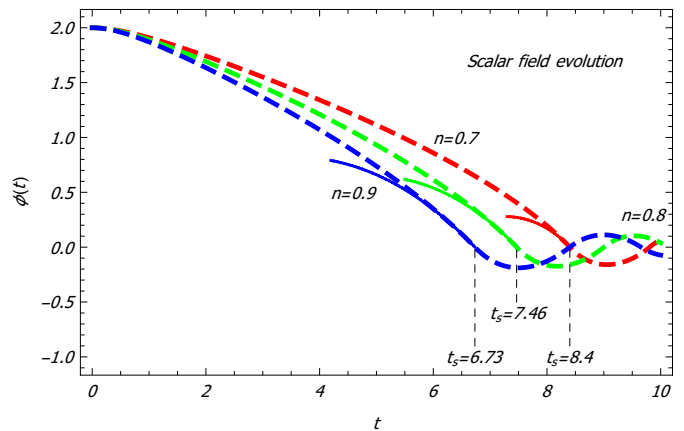}}}
\caption{Numerical (dashed line) and analytical (continuous line) time evolution of the scalar field, for $n=0.7, 0.8, 0.9$. The two solutions for each $n$ are consistent close to each singularity.}
\label{fig5}
\end{figure}

\begin{figure}[!h]
\centering
\vspace{0.3cm}\rotatebox{0}{\vspace{0cm}\hspace{0cm}\resizebox{0.49\textwidth}{!}{\includegraphics{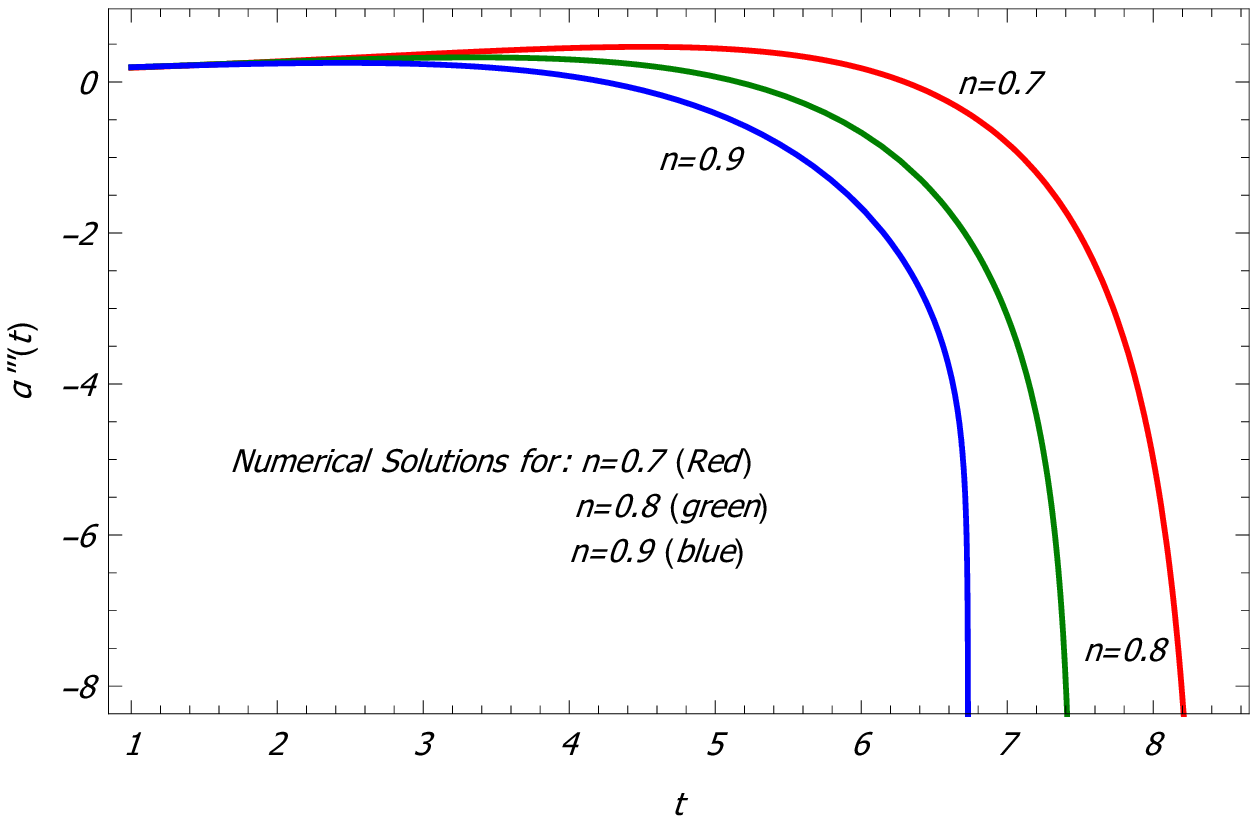}}}
\caption{Numerical solutions of the third time derivative of the scale factor for $n=0.7, 0.8, 0.9$. Notice the divergence at the time of the singularity when the scalar field vanishes ($t_s=8.4$ for $n=0.7$, $t_s=7.46$ for $n=0.8$, $t_s=6.73$ for $n=0.9$).}
\label{fig6}
\end{figure}

\begin{figure}[!h]
\centering
\vspace{0.3cm}\rotatebox{0}{\vspace{0cm}\hspace{0cm}\resizebox{0.49\textwidth}{!}{\includegraphics{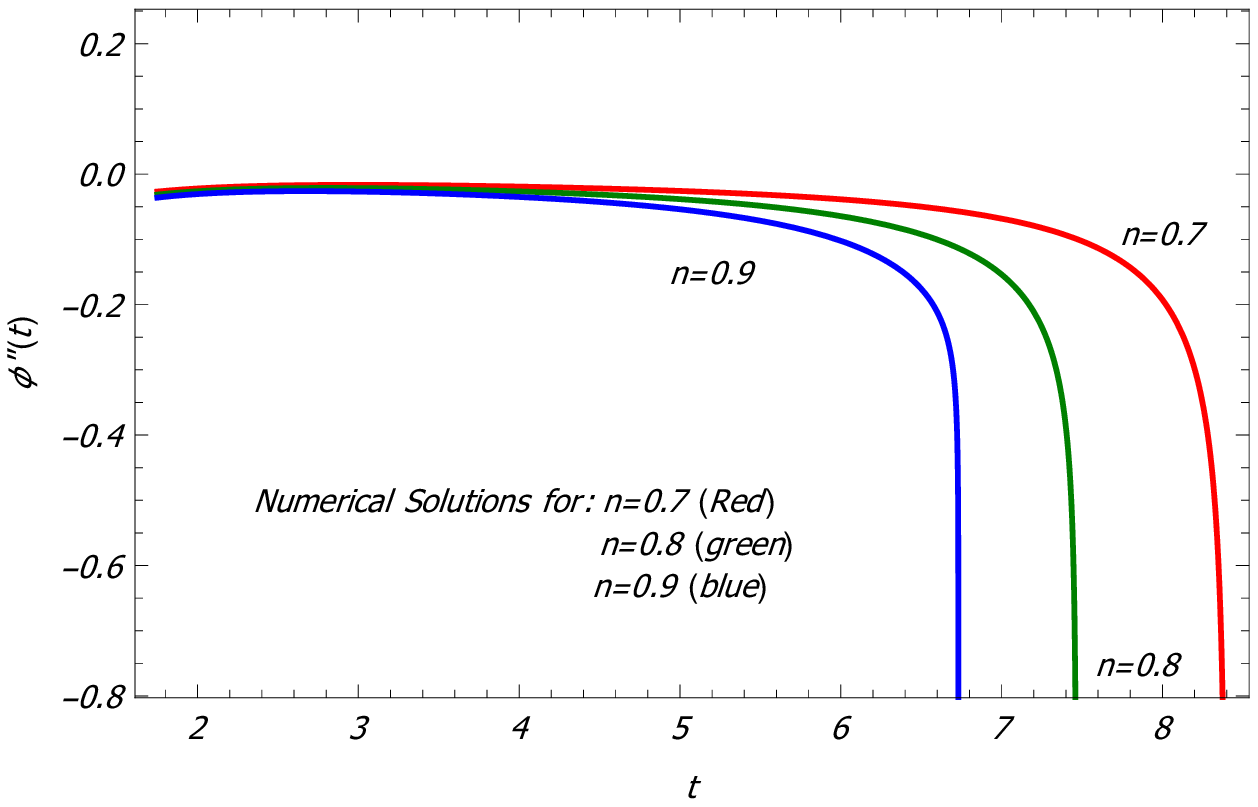}}}
\caption{Numerical solutions of the second time derivative of the scalar field for $n=0.7, 0.8, 0.9$. Notice the divergence at the time of the singularity when the scalar field vanishes ($t_s=8.4$ for $n=0.7$, $t_s=7.46$ for $n=0.8$, $t_s=6.73$ for $n=0.9$).}
\label{fig7}
\end{figure}

Substituting the expressions (\ref{scalefactor1}), (\ref{scalarfield}), (\ref{potential}) for $a(t), \phi(t)$ and $V(\phi)$ in the dynamical eqs. (\ref{barrow1}) and (\ref{barrow3}), it is straightforward to calculate the relations between the coefficients $c,d,f,h$. The form of the relations between the evaluated expansion coefficients, is shown in the Appendix, and has been verified by numerical solution of the dynamical equations.  

The additional linear and quadratic terms in $(t_{s}-t)$, in the expression of the scale factor (\ref{scalefactor1}), play an important role in the estimation of the Hubble parameter and its derivative as the singularity is aproached. 

The relations between these coefficients can lead to relations between the Hubble parameter and its derivative close to the singularity, which in turn correspond to observational predictions that may be used to identify the presence of these singularities in angular diameter of luminosity distance data. For example, the coeficients $b$ and $c$ are related as (see Appendix eq. (\ref{c-b})),

\be 
c=-\frac{b^2}{a_{s}}.
\ee

\noindent Using this relation, it is easy to show that
\be \label{dotH}
\dot H=-3H^2
\ee
\noindent (see proof in Appendix). This result constitutes an observationally testable prediction of this class of models, which can be used to search for such singularities in our past light cone.

\subsection{Numerical analysis}

It is straightforward to verify numerically the derived power law dependence of the scale factor and scalar field as the singularity is approached. We thus solve the rescaled, with the present day Hubble parameter $H_{0}$ (setting $H=\bar H H_{0}$, $t=\bar t/ H_{0}$, $V=\bar V H^{2}_{0}$ ), coupled system of the cosmological dynamical equations for the scale factor and for the scalar field (\ref{barrow2}) and (\ref{barrow3}). We assume initial conditions at early times ($t\ll t_{0}$) when the scalar field is assumed frozen at $\phi(t_{i})=\phi_{i}$ and $\dot \phi(t_{i})=0$ due to cosmic friction \cite{Caldwell:2005tm, Scherrer:2007pu}. At that time the initial conditions for the scale factor are well approximated by
\be
a(t_{i})=\exp \left [\sqrt {\frac{V(\phi_{i})}{3}} t_{i} \right],
\ee

\be
\dot a(t_{i}))=\sqrt {\frac{V(\phi_{i})}{3}} \left [ \exp \sqrt {\frac{V(\phi_{i})}{3}} t_{i} \right]
\ee
Taking the logarithm of the numerical solution corresponding to the third derivative of the scale factor (\ref{scalefactor1}) and to the second derivative of the scalar field (\ref{scalarfield}), we obtain Fig.\ref{fig:fig2} and Fig.\ref{fig:fig3}, which show these logarithms as functions of $t_s-t$ close to the singularity (continous lines). On these lines we superpose the corresponding analytic expansions (eqs (\ref{scalefactor1}) and (\ref{scalarfield}), dashed lines) which, close to the singularity, may be written as

\be \label{loga}
\log [|\dddot a|]=\log [|d|q(q-1)(q-2)]+(q-3) \log [(t_{s}-t)] 
\ee

\noindent and

\be \label{logf}
\log [|\ddot \phi|]=\log [|h|r(r-1)]+(r-2) \log [(t_{s}-t)]. 
\ee

In the plots of eqs (\ref{loga}), (\ref{logf}) (dashed lines) we have used the predicted values of the exponents (eqs (\ref{r}) and (\ref{q1})) and the analytically predicted values for the coeficients $d$ and $h$ shown in the Appendix. We underline the good agreement in the slopes of the analytically predicted curves and the corresponding numerical results, which confirm the validity of the power law ansatz (\ref{scalefactor1}), (\ref{scalarfield}), and the values of the corresponding exponents (\ref{r}), (\ref{q1})). 

We have also verified this agreement by obtaining the best fit slopes of the numerical solutions of Fig.\ref{fig:fig2}, Fig.\ref{fig:fig3} deriving the numerically predicted values of the exponents $q$ and $r$. These numerical best fit values, along with the corresponding analytical predictions, are shown in Table \ref{TabII} for $n=0.5$ and $n=0.7$, indicating good agreement between the analytical and numerical values of the exponents. 

In Fig.\ref{fig4}, Fig.\ref{fig5} we show the time evolution (numerical and analytical) of the scale factor and the scalar field respectively. The two curves, for each $n$, are consistent close to each singularity. In Fig.\ref{fig6} and Fig.\ref{fig7} we demonstrate numerically the divergence of the third derivative of the scale factor and of the second derivative of the scalar field. The divergence occurs at the time of the singularity when the scalar field vanishes \ie $\phi=0$.

\begin{table}[h!]
\centering
\scalebox{1.15}{\begin{tabular}{c|c|c|c|c|}
\cline{2-5}
& \multicolumn{2}{c|}{Numerical} &\multicolumn{2}{c|}{Analytical}  \\ \hline  
\multicolumn{1}{|c|}{$n$}             &  $r$            & $q$             & $r=n+1$      & $q=n+2$   \\ \hline \hline
\multicolumn{1}{|c|}{$0.5$}  & $1.5 \pm 0.0003$ &      $2.51 \pm 0.0007$   &   $1.5$  &    $2.5$    \\ \hline
\multicolumn{1}{|c|}{$0.7$}  & $1.7 \pm 0.002$ &    $2.71 \pm 0.004$     &     $1.7$  &       $2.7$    \\ \hline
\end{tabular}}
\caption{Numerical and analytical values of the power exponents $r, q$. Clearly, there is consistency between numerical results and analytical expectations.}
\label{TabII}
\end{table}

\subsection{Evolution with a perfect fluid}

In the presence of a perfect fluid, the action of the theory is obtained from the generalized action (\ref{action1}) with $F(\phi)=1$ as

\be \label{action3}
\mathcal{S}=\int \left [\frac{1}{2}R+\frac{1}{2}g^{\mu \nu} \phi_{;\mu}  \phi_{;\nu}-V(\phi)+\mathcal{L}_{(fluid)} \right ]\sqrt{-g}d^{4}x.
\ee

The corresponding dynamical equations are

\be \label{barrow1matter}
3H^2=\frac{3\Omega_{0,m}}{a^{3}}+ \frac{1}{2}\dot{\phi}^2+V(\phi)
\ee

\be \label{barrow2matter}
\ddot{\phi}=-3H\dot{\phi}-An|\phi|^{n-1}  \Theta(\phi)
\ee

\be \label{barrow3matter}
2\dot H=-\frac{3\Omega_{0,m}}{a^{3}}-\dot \phi^2
\ee

\noindent with $\rho_{m}=\frac{\rho_{0m}}{a^{3}}=\frac{3\Omega_{0,m}}{a^{3}}$ and $\Omega_{0,m}=0.3$. The scale factor (\ref{scalefactor1}), in the presence of a perfect fluid is now assumed to be of the form

\be \label{generscalefactor}
a(t)=1+(a_s-1)\left(\frac{t}{t_s}\right)^m+b(t_{s}-t) + c(t_{s}-t)^2 +d(t_{s}-t)^q
\ee

\noindent where $m=\frac{2}{3(1+w)}$ and $w$ the state parameter.
\noindent As in the case of the previous section, from the dynamical equations (\ref{barrow1matter}), (\ref{barrow3matter}), $H, \dot H, \dot \phi$ still remain finite. Also in eq. (\ref{barrow2matter}) there is a divergence of the term $\phi^{n-1}$ for $0<n<1$ and $\ddot \phi \to \infty$ as $\phi \to 0$. The third derivative of the scale factor $\dddot a$ also diverges due to the divergence of $\ddot H$ (differentiation of eq. (\ref{barrow3})). Thus, the constraints for $q,r$ are the same as in the absence of the fluid (section II.1), \ie $2<q<3$ and $1<r<2$ respectively.

Following the steps of section II.1, we rediscover the same values for the exponents \ie  eqs (\ref{r}) and (\ref{q1}) which imply similar behaviour close to the singularity.

The relations among the expansion coefficients $c,d,f,h$, are shown in the Appendix, and have been verified by numerical solution of the dynamical equations, as in the absence of the fluid (see Appendix). For $\rho_{0m}=0$ all coefficients reduce to those of the no fluid case.

An interesting result arises from the derivation of the relation between the coefficients $b, c$. The relation between $b, c$  in the presence of a fluid is of the form (see Appendix eq. (\ref{cmatter})),

\be \label{cmatter1}
c=\frac{\rho_{0,m}}{4a^{2}_{s}}-\frac{1}{2}(a_{s}-1)m(m-1)-\frac{[(a_{s}-1)m-b]^{2}}{a_{s}},
\ee

\noindent Thus, close to the singularity we obtain 

\be \label{dotHmatterred}
\dot H=\frac{3}{2}\Omega_{0,m} (1+z_{s})^{3}-3H^{2}
\ee
where $z_s$ is the redshift at the time of the singularity. Clearly eq. (\ref{dotHmatterred}) reduces to eq. (\ref{dotH}) for $\rho_{0,m}=0$ (see proof in Appendix). This result may be used as observational signature of such singularities in this class of models.

\section{Sudden Future Singularities in Scalar-Tensor Quintessence Models}
\label{sec:Section 3}

\subsection{Evolution without a perfect fluid}

We now consider now scalar-tensor quintessence models without the presence of a perfect fluid. The action of the theory is the generalized action (\ref{action1}), where $\mathcal{L}_{(fluid)}$ is ignored. Therefore, it has the form

\be
\mathcal{S}=\int \left [\frac{1}{2}F(\phi)R+\frac{1}{2}g^{\mu \nu} \phi_{;\mu}  \phi_{;\nu}-V(\phi) \right ]\sqrt{-g}d^{4}x
\ee

We assume a nonminimal coupling linear in the scalar field $F(\phi)=1-\lambda \phi$ even though our results about the type of the singularity in this class of models is unaffected by the particular choice of the nonminimal coupling. The dynamical equations are of the form

\be \label{periv1}
3FH^{2}=\frac{\dot \phi^2}{2}+V-3H\dot{F}
\ee

\be \label{periv2}
\ddot \phi+3H\dot \phi-3F_{\phi}\bigg(\frac{\ddot a}{a}+H^2\bigg)+An|\phi|^{(n-1)} \Theta(\phi)=0
\ee

\be \label{periv3}
-2F\bigg(\frac{\ddot a}{a}-H^{2}\bigg)=\dot \phi^{2}+\ddot F-H\dot F,
\ee

\noindent where $F_{\phi}=\frac{d}{d\phi} F$. From eq. (\ref{periv1}), it is clear that $H, \dot \phi, F, \dot F$ all remain finite when $\phi \to 0$ ($t\to t_{s}$). However, in eq. (\ref{periv2}) there is a divergence of the term $V_{\phi}$ for $0<n<1$ and $\ddot \phi \to \infty$ as $\phi \to 0$. This means that $\ddot F\to \infty$ because of the generation of the second derivative of $\phi$ that leads to a divergence of $\ddot a$ in eq. (\ref{periv3}). The effective dark energy density and pressure take the form  \cite{Capozziello:2007iu, Lykkas:2015kls}
\be \label{rDE}
\rho_{DE}=\frac{\dot \phi^2}{2}+V-3FH^{2}-3H\dot{F}
\ee

\be \label{pDE}
p_{DE}=\frac{\dot \phi^2}{2}-V-(2\dot H-3H^{2})F+\ddot F+2H\dot F.
\ee
Thus $\rho_{DE}$ remains finite in eq. (\ref{rDE}), while $p_{DE} \to \pm \infty$ in eq. (\ref{pDE}). Clearly, an SFS singularity (Table \ref{TabI}, see also \cite{Dabrowski:2007ci}) is expected to occur in scalar-tensor quintessence models, as opposed to the GSFS singularity in the corresponding quintessence models. This result will be verified quantitatively in what follows.

Using the ansatz (\ref{scalefactor1}), (\ref{scalarfield}) in the dynamical eq. (\ref{periv3}) we find that the dominant terms close to the singularity are

\be \label{intermediate1sqm}
\mathcal{B}_{1}(t_{s}-t)^{q-2}=\mathcal{B}_{2}(t_{s}-t)^{r-2}
\ee
\noindent where the $\mathcal{B}_{1}, \mathcal{B}_{2}$ are constants, which depend on the coefficient $d, h$ and the $\lambda$ constant, and are shown in the Appendix.
It immediately follows from  eq. (\ref{intermediate1sqm}) that
\be \label{q2}
q=r
\ee
Similarly, substituting the ansatz (\ref{scalefactor1}), (\ref{scalarfield}) in eq. (\ref{periv2}) we find that the dominant terms close to the singularity obey the equation

\be \label{intermediate2sqm}
\mathcal{B'}_{1}(t_{s}-t)^{r-2}=\mathcal{B'}_{2}(t_{s}-t)^{n-1}
\ee
\noindent where the $\mathcal{B'}_{1}, \mathcal{B'}_{2}$ are constants, which depend on the coefficient $f$ and the constants  $A, \lambda$  as shown in the Appendix.
Equating the exponents of the divergent terms we find
\be \label{r2}
r=n+1,
\ee
\noindent which  leads to
\be \label{q-n2}
q=n+1.
\ee

\begin{figure}[!h]
\centering
\vspace{0.3cm}\rotatebox{0}{\vspace{0cm}\hspace{0cm}\resizebox{0.48\textwidth}{!}{\includegraphics{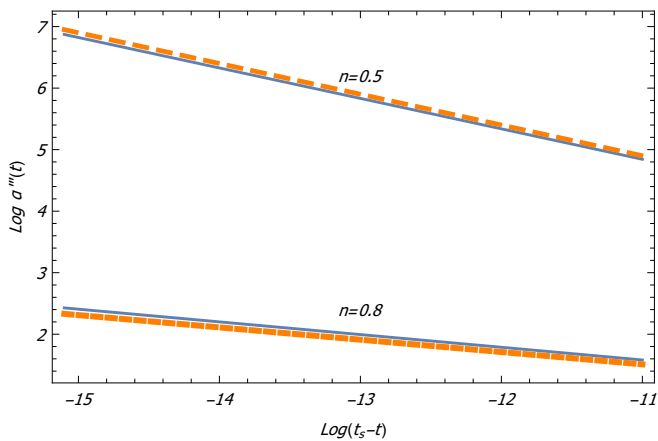}}}
\caption{Numerical verification of the $q$-exponent for $n=0.5$ and $n=0.8$. The orange dashed line, denotes the analytical, while the blue line denotes the numerical solution. As expected the slopes for each n are identical, while the small difference is due to the coefficients.}
\label{fig:fig8}
\end{figure}

\begin{figure}[!h]
\centering
\vspace{0.3cm}\rotatebox{0}{\vspace{0cm}\hspace{0cm}\resizebox{0.48\textwidth}{!}{\includegraphics{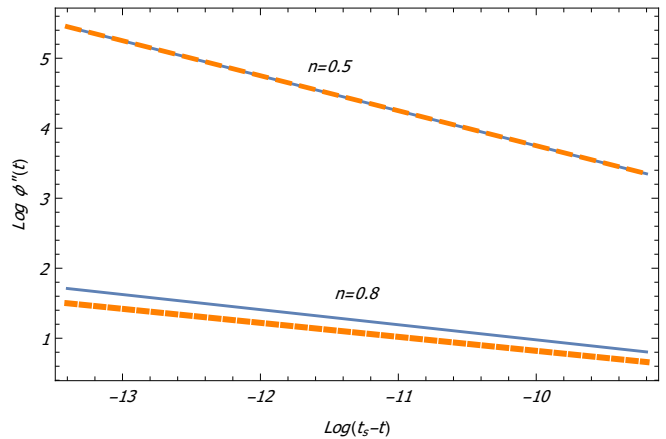}}}
\caption{Numerical verification of the $r$-exponent for $n=0.5$ and $n=0.8$. The orange dashed line, denotes the analytical, while the blue line denotes the numerical solution. As expected the slopes for each n are identical, while the small difference is due to the coefficients.}
\label{fig:fig9}
\end{figure}

The results (\ref{r2}) and (\ref{q-n2}) are consistent with the above qualitative discussion for the expected strength of the singularity. Thus in the case of the scalar-tensor theory we have a stronger singularity at $t_{s}$, compared to the singularity that occurs in quintessence models. This is a general result, valid not only for the coupling constant of the form $F=1-\lambda \phi$ but also for other forms of $F(\phi)$ (\eg $F\sim \phi^r$), because the second derivative of $F$
with respect to time, in the dynamical equations, will always generate a second derivative of $\phi$ with divergence, leading to a divergence of $\ddot a$. 

\begin{figure}[!h]
\centering
\vspace{0.3cm}\rotatebox{0}{\vspace{0cm}\hspace{0cm}\resizebox{0.48\textwidth}{!}{\includegraphics{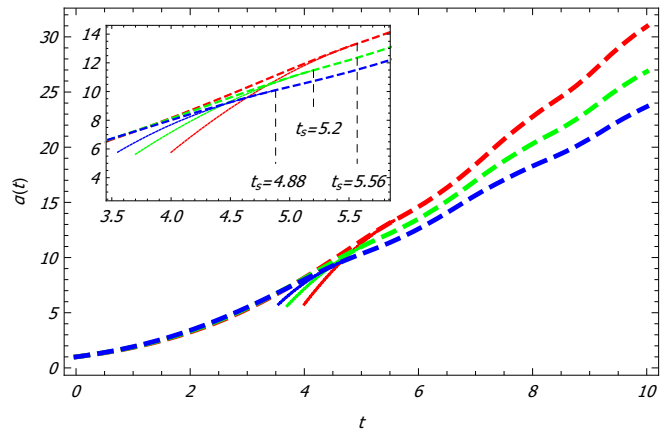}}}
\caption{Plot of numerical (dashed) and analytical (line) time evolution of the scale factor, for $n=0.7$ (red), $0.8$ (green), and $0.9$ (blue). The two solutions for each $n$ are consistent close to each singularity.}
\label{fig10}
\end{figure}

\begin{figure}[!h]
\centering
\vspace{0.3cm}\rotatebox{0}{\vspace{0cm}\hspace{0cm}\resizebox{0.48\textwidth}{!}{\includegraphics{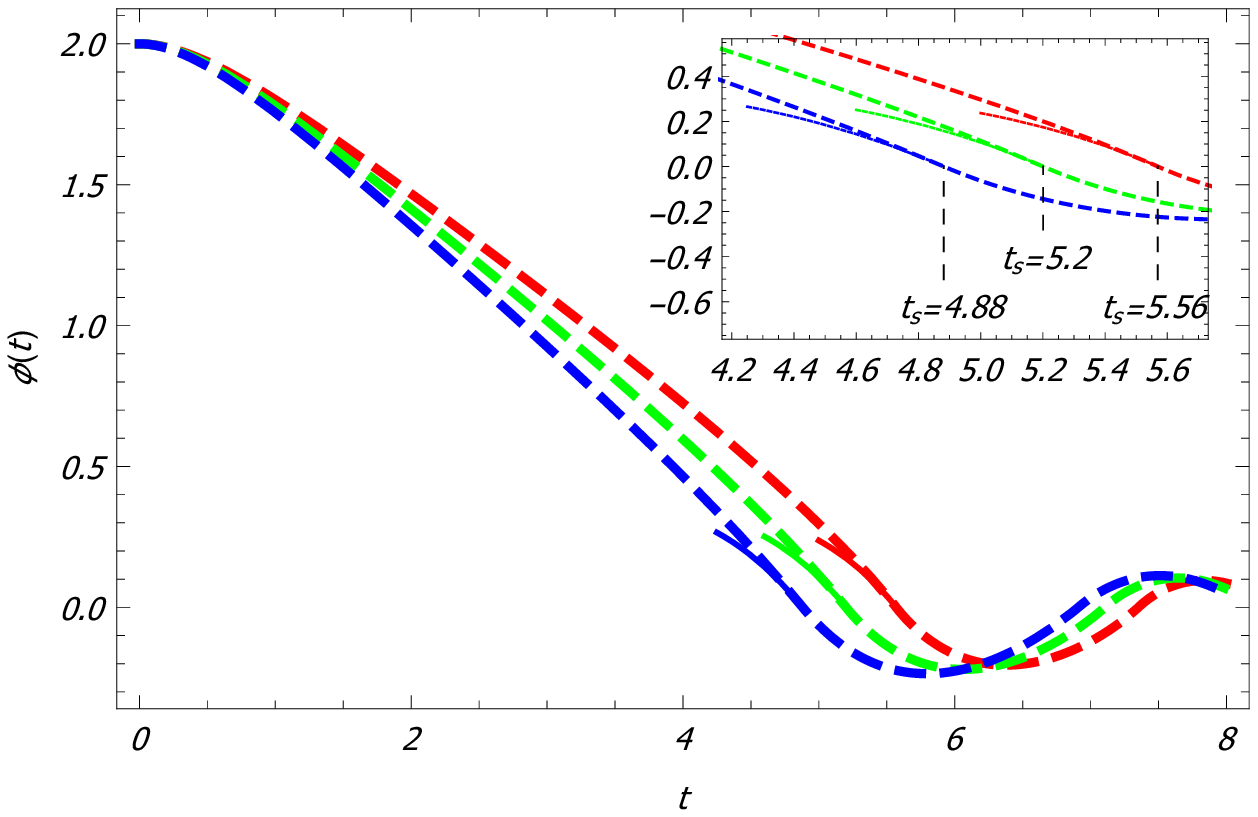}}}
\caption{Plot of numerical (dashed) and analytical (line) time evolution of the scalar field, for for $n=0.7$ (red), $0.8$ (green), and $0.9$ (blue). The two solutions for each $n$ are consistent close to each singularity.}
\label{fig11}
\end{figure}

\begin{figure}[!h]
\centering
\vspace{0.3cm}\rotatebox{0}{\vspace{0cm}\hspace{0cm}\resizebox{0.48\textwidth}{!}{\includegraphics{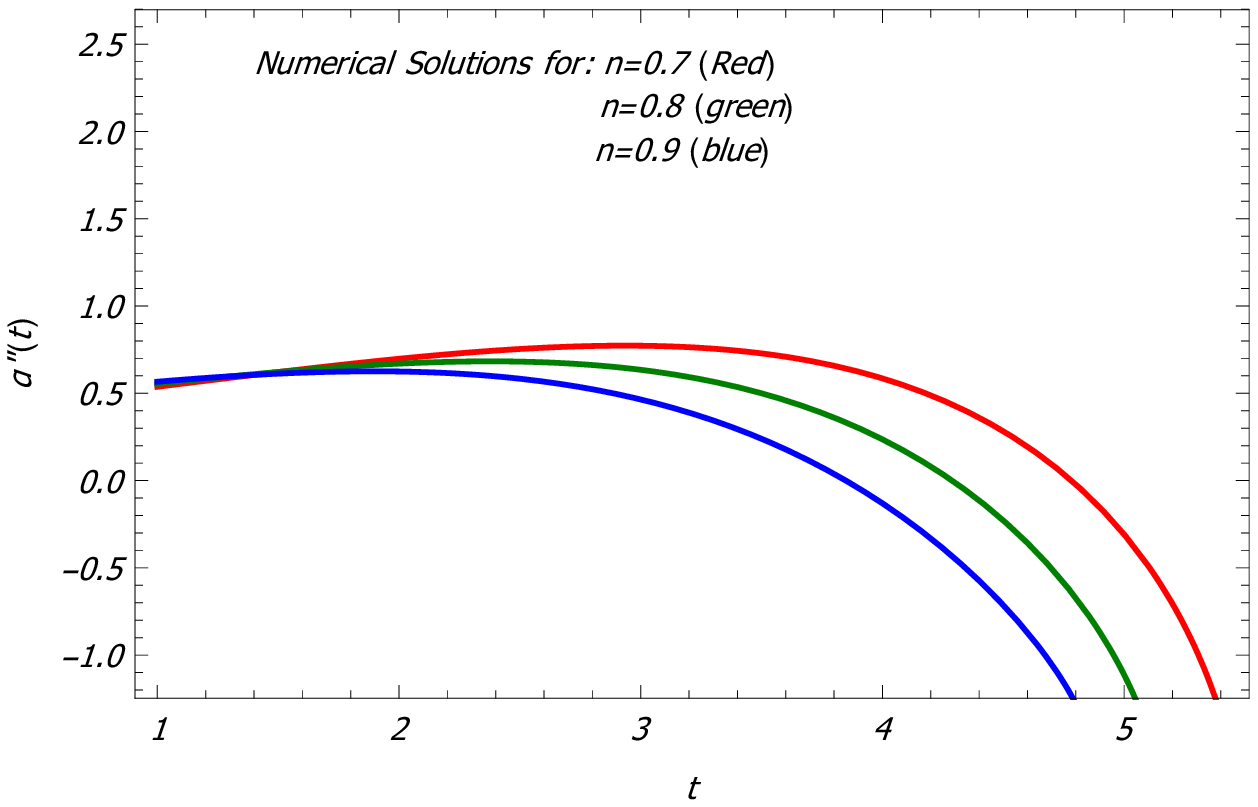}}}
\caption{Numerical solutions of the second time derivative of the scale factor for $n=0.7, 0.8, 0.9$. Notice the divergence at the time of the singularity when the scalar field vanishes ($t_s=5.56$ for $n=0.7$, $t_s=5.2$ for $n=0.8$, $t_s=4.88$ for $n=0.9$).}
\label{fig12}
\end{figure}

\begin{figure}[!h]
\centering
\vspace{0.3cm}\rotatebox{0}{\vspace{0cm}\hspace{0cm}\resizebox{0.48\textwidth}{!}{\includegraphics{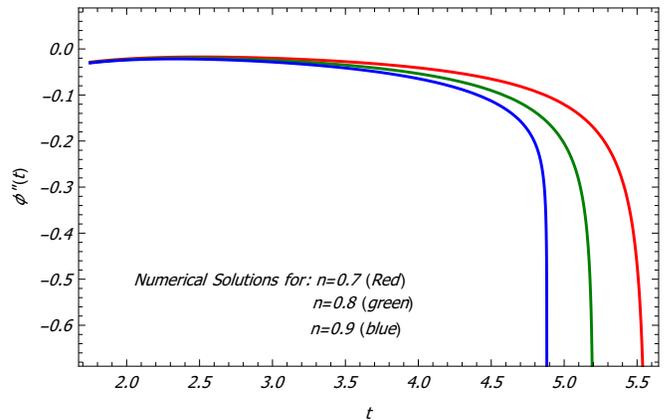}}}
\caption{Numerical solutions of the second time derivative of the scalar field for $n=0.7, 0.8, 0.9$. Notice the divergence at the time of the singularity when the scalar field vanishes ($t_s=5.56$ for $n=0.7$, $t_s=5.2$ for $n=0.8$, $t_s=4.88$ for $n=0.9$).}
\label{fig13}
\end{figure}

Using eqs (\ref{periv1}), (\ref{intermediate1sqm}), (\ref{q2}), (\ref{intermediate2sqm}) and (\ref{r2}), we calculate relations among the coefficients $c,d,f,h$. The form of these relations, is shown in the Appendix, and has been verified by numerical solution of the dynamical equations. Notice that all coefficients, except $d$, reduce to those of section II.1 for $\lambda=0$. \footnote{The coefficient $d$ differs in scalar-quintessence since the divergence occurs in the second, instead of the third derivative of the scale factor.} 

\subsection{Numerical analysis}

We now solve the rescaled coupled system of the cosmological dynamical equations for the scale factor and for the scalar field (\ref{periv2}) and (\ref{periv3}), using the present day Hubble parameter $H_{0}$ (setting $H=\bar H H_{0}$, $t=\bar t/ H_{0}$, $V=\bar V H^{2}_{0}$). We assume initial conditions at early times ($t\ll t_{0}$) when the scalar field is assumed frozen at $\phi(t_{i})=\phi_{i}$ and $\dot \phi(t_{i})=0$ due to cosmic friction in the context of thawing \cite{Caldwell:2005tm, Scherrer:2007pu} scalar-tensor quintessence \cite{Perivolaropoulos:2005yv, Nesseris:2006hp, Nesseris:2006er}. At that time the initial conditions for the scale factor are 

\be
a(t_{i})=\exp \left [\sqrt {\frac{V(\phi_{i})}{3F_{i}}} t_{i} \right],
\ee

\be
\dot a(t_{i}))=\exp \left [ \sqrt {\frac{V(\phi_{i})}{3F_{i}}} t_{i} \right] \sqrt {\frac{V(\phi_{i})}{3F_{i}}}, 
\ee

\noindent where $F_{i}=1-\lambda \phi_{i}$.

Taking the logarithm of the second derivative of the scale factor (\ref{scalefactor1}) and of the scalar field (\ref{scalarfield}), we obtain

\be \label{logaa}
\log [|\ddot a|]=\log [|d|q(q-1)]+(q-2) \log [(t_{s}-t)] 
\ee

\noindent and

\be \label{logff}
\log [|\ddot \phi|]=\log [|h|r(r-1)]+(r-2) \log [(t_{s}-t)] 
\ee

The numerical verification of the validity of  eqs (\ref{r2}), (\ref{q-n2}) has been performed similarly to the case of minimally coupled quintessence. In Fig.\ref{fig:fig8} and Fig.\ref{fig:fig9} we show the analytical and numerical solutions, for the logarithm of the diverging terms of the scale factor and the scalar field respectively, as $t\to t_{s}$ from below. The $\log$-plots of the diverging terms of $\ddot a$ and $\ddot \phi$ are straight lines, indicating a power law behaviour with best fit slopes as shown in Table \ref{TabIII}, in good agreement with the analytical expansion expectations (eqs (\ref{r2}), (\ref{q-n2}). In Figs.\ref{fig10}, \ref{fig11} we show the time evolution (numerical and analytical) of the scale factor and the scalar field respectively. The two curves, for each $n$, are consistent close to each singularity. In Figs.\ref{fig12}, \ref{fig13} we demonstrate numerically the divergence of the second derivarive of the scale factor and of the scalar field. As expected, the divergence occurs at the time of the singularity when the scalar field vanishes.

\begin{table}[h!]
\centering
\scalebox{1.15}{\begin{tabular}{c|c|c|c|c|}
\cline{2-5}
& \multicolumn{2}{c|}{Numerical} &\multicolumn{2}{c|}{Analytical}  \\ \hline  
\multicolumn{1}{|c|}{$n$}             &  $r$            & $q$             & $r=n+1$      & $q=n+1$   \\ \hline \hline
\multicolumn{1}{|c|}{$0.5$}  & $1.5 \pm 0.0003$ &    $1.49 \pm 0.0002$     &     $1.5$  &       $1.5$    \\ \hline
\multicolumn{1}{|c|}{$0.8$}  & $1.8 \pm 0.03$ &      $1.8 \pm 0.006$   &   $1.8$  &    $1.8$    \\ \hline
\end{tabular}}
\caption{Numerical and analytical values of the power-laws $r, q$.  Clearly, there is consistency between numerical results and analytical expectations.}
\label{TabIII}
\end{table}

Using eqs (\ref{logaa}), (\ref{logff}), it is straightforward to obtain numerically the values of the parameters $h$ of the scalar field, as well as $d$ of the scale factor, and compare with their analytically obtained values shown in the Appendix.

The quadratic term of $(t_{s}-t)$, in the expression of the scale factor (\ref{scalefactor1}), is now subdominant as the second derivarive of the scale factor diverges. The only additional term of $(t_{s}-t)$ that can play an important role in the estimation of the Hubble parameter, is the linear term. Clearly, for the first derivative of (\ref{scalefactor1}), as $t\to t_{s}$ from below, the linear term dominates over all other terms, while the quadratic term is subdominant in the second derivative in the divergence of the $q$-term. Thus, in the case of the scalar-tensor quintessence models $H$ remain finite and dominated by the term $b(t_{s}-t)$, while $\dot H \to \infty$ as $t\to t_{s}$.

\subsection{Evolution with a perfect fluid}

In the presence of a perfect fluid, the action is now the generalized action (\ref{action1}). The scale factor and the scalar field are of the form (\ref{generscalefactor}) and (\ref{scalarfield}) respectively. The dynamical equations in the presence of a relativistic fluid become

\be \label{perivol1matter}
3FH^{2}=\frac{3\Omega_{0,m}}{a^{3}}+\frac{\dot \phi^2}{2}+V-3H\dot{F}
\ee

\be \label{perivol2matter}
\ddot \phi+3H\dot \phi-3F_{\phi}\bigg(\frac{\ddot a}{a}+H^2\bigg)+V_{\phi}=0
\ee

\be \label{perivol3matter}
-2F\bigg(\frac{\ddot a}{a}-H^{2}\bigg)=\frac{3\Omega_{0,m}}{a^{3}}+\dot \phi^{2}+\ddot F-H\dot F
\ee

The constraints for $r, q$ as $t\to t_{s}$ from below, are the same as in the absence of the fluid \ie $1<r<2$ and $1<q<2$, and following the steps of the section III.1 we obtain

\be \label{q3}
q=r,
\ee

\be \label{r3}
r=n+1,
\ee

and according to eq. (\ref{q3}) 

\be
q=n+1.
\ee

\noindent \ie eqs (\ref{q2}), (\ref{r2}) and (\ref{q-n2}) respectively.
Finally, the form of the evaluated expansion coefficients $c,d,f,h$ is shown in the Appendix, and has been verified by numerical solution of the dynamical equations (see Appendix). As expected, for $\rho_{0m}=0$, all coefficients reduce to the ones of the no fluid case. 

\section{Conclusion-Discussion}
\label{sec:Section 4}

We have derived analytically and numerically the cosmological solution close to a future-time singularity for both quintessence and scalar-tensor quintessence models. For quintessence, we have shown that there is a divergence of $\dddot a$ and a GSFS singularity occurs ($a_{s}, \rho_{s}, p_{s}$ remain finite but $\dot p \to \infty)$ , while in the case of scalar-tensor quintessence models there is a divergence of $\ddot a$ and an SFS singularity occurs ($a_{s}, \rho_{s}$ remain finite but $p_{s}\to \infty$, $\dot p \to \infty)$. Importing a perfect fluid in the dynamical equations, in both cases, we have shown that this result is still valid in our cosmological solution.

These are the simplest non-exotic physical models where GSFS and SFS singularities naturally arise. In the case of scalar-tensor quintessence models, there is a divergence of the scalar curvature $R=6\left (\frac{\ddot {a}}{a}+\frac{\dot a^{2}}{a^{2}} \right ) \to \infty$ because of the divergence of the second derivative of the scale factor. Thus, a stronger singularity occurs in this class of models. Such divergence of the scalar curvature is not present in the simple quintessence case.

We have also shown the important role of the additional linear and quadratic terms of $t_{s}-t$ in the form of the scale factor as $t\to t_{s}$. However, in the scalar-tensor case the quadratic term becomes subdominant close to the singularity.

We have derived explicitly the relations between the coefficients of the linear, quadratic and diverging terms of the scale factor and the scalar field. We have shown that all coefficients of the fluid case (quintessence and scalar-tensor quintessence), reduce to those of the no fluid case for $\rho_{0m}=0$, and all coefficients (except coefficient $d$) of the scalar-tensor models reduce to those of the simple quintessence, in the special case $\lambda=0$ \ie $F=1$. Moreover, for quintessence models, we derived relations of the Hubble parameter, $\dot H=-3H^{2}$ (for the no fluid case) and $\dot H=\frac{3}{2}\Omega_{0,m} (1+z_{s})^{3}-3H^{2}$ (for the fluid case), close to the singularity. These relations may be used as observational signatures of such singularities in this class of models.

Interesting extensions of the present analysis include the study of the strength of these singularities in other modified gravity models \eg string-inspired gravity, Gauss-Bonnet gravity etc \cite{Nojiri:2008fk, Nojiri:2006ri} and the search for signatures of such singularities in cosmological luminosity distance and angular diameter distance data.

{\bf Numerical Analysis:} The Mathematica file that led to the production of the figures may be downloaded from \href{http://leandros.physics.uoi.gr/quint-singularities/math-quint.zip}{here}.


\appendix*
\section*{appendix}

\subsection*{Relations among the expansion coefficients}

\subsubsection*{Quintessence without matter}

Substituting the expressions (\ref{scalefactor1}), (\ref{scalarfield}), (\ref{potential}) for $a(t), \phi(t)$ and $V(\phi)$ in the dynamical eqs. (\ref{barrow1}) and (\ref{barrow3}), it is straightforward to obtain relations among the expansion coefficients as 

\be
f=\frac{b}{a_{s}}\sqrt{6}
\ee

\be \label{c-b}
c=-\frac{b^2}{a_{s}}.
\ee

\be \label{h}
h=-\frac{Af^{n-1}}{n+1}
\ee

\be
d=\frac{Ab\sqrt{6}f^{n-1}}{(n+1)(n+2)}.
\ee

Also eq. (\ref{intermediate}) may be written explicitly as

\be \nn
hr(r-1)(t_{s}-t)^{r-2}=-Anf^{n-1}(t_{s}-t)^{n-1}
\ee

Thus, the constants $\mathcal{A}_{1}, \mathcal{A}_{2}$ are

\be
\mathcal{A}_{1}=hr(r-1)
\ee

\be
\mathcal{A}_{2}=-Anf^{n-1}
\ee

Similarly eq  (\ref{intermediate1}) may be written explicity as

\be \nn
\frac{dq(q-1)(q-2)}{a_{s}}(t_{s}-t)^{q-3}=-fhr(r-1)(t_{s}-t)^{r-2}
\ee

Thus, the constants $\mathcal{A'}_{1}, \mathcal{A'}_{2}$ are of the form

\be
\mathcal{A'}_{1}=\frac{dq(q-1)(q-2)}{a_{s}}
\ee

\be
\mathcal{A'}_{2}=-fhr(r-1)
\ee

\subsubsection*{Quintessence with matter}

As in the previous case from the dynamical equations eq. (\ref{barrow1matter}, \ref{barrow3matter}) we find the corresponding expansion coefficients 

\be \label{fmatter}
f=\left [6\frac{\left ((a_{s}-1)m-b\right )^{2}}{a^{2}_{s}}-2\frac{\rho_{0,m}}{a^{3}_{s}} \right ]^{1/2}, 
\ee

\be \label{cmatter}
c=\frac{\rho_{0,m}}{4a^{2}_{s}}-\frac{1}{2}(a_{s}-1)m(m-1)-\frac{[(a_{s}-1)m-b]^{2}}{a_{s}},
\ee

\be
h=-\frac{Af^{n-1}}{n+1}
\ee

\be
d=\frac{Af^{n-1}}{(n+1)(n+2)} \sqrt{6[(a_{s}-1)m-b]^{2}-2\frac{\rho_{0,m}}{a_{s}}}.
\ee

For $m=\rho_{0m}=0$ t all coefficients reduce to the previous ones of the no fluid case as expected.

\subsubsection*{Scalar-tensor quintessence without matter}

In this case the dynamical equations lead to the following relations among the expansion coefficients

\be \label{result}
f=-\frac{3\lambda b}{a_{s}} \pm \frac{\sqrt{3} \sqrt{b^{2}(2+3\lambda^2)}}{a_{s}}
\ee

\be
d=\frac{1}{2}\lambda a_{s}h
\ee

\be
c=-\frac{b^2}{a_{s}}+\frac{5}{4}\lambda bf.
\ee

\be
h=-\frac{Af^{n-1}}{(n+1)\big(1+\frac{3}{2}\lambda^{2}\big)}.
\ee

We notice that all coefficients except $d$, reduce to those of section II.1 for $\lambda=0$. The reason that the coefficient $d$ differs in scalar-quintessence is because in this case the divergence occurs in the second, instead of the third derivative of the scale factor.

Eq. (\ref{intermediate1sqm}) is written explicitly, keeping only the dominant terms

\be \nn
2\frac{dq(q-1)}{a_{s}}(t_{s}-t)^{q-2}=\lambda hr(r-1)(t_{s}-t)^{r-2}
\ee

Thus, the constants $\mathcal{B}_{1}, \mathcal{B}_{2}$ are

\be
\mathcal{B}_{1}=2\frac{dq(q-1)}{a_{s}}
\ee

\be
\mathcal{B}_{2}=\lambda hr(r-1)
\ee

Similarly, eq. (\ref{intermediate2sqm}) is written explicitly, keeping only the dominant terms

\be \nn
\left (\frac{3}{2}\lambda^{2}+1\right )r(r-1)(t_{s}-t)^{r-2}=-Anf^{n-1}(t_{s}-t)^{n-1}
\ee

Thus, the constants $\mathcal{B'}_{1}, \mathcal{B'}_{2}$ are

\be
\mathcal{B'}_{1}=\left (\frac{3}{2}\lambda^{2}+1\right )r(r-1)
\ee

\be
\mathcal{B'}_{2}=-Anf^{n-1}
\ee

\subsubsection*{Scalar-tensor quintessence with matter}

As in the previous cases we use the relevant dynamical equation which in this case is eq. (\ref{perivol1matter}) to obtain the relations among the expansion coefficients as

\be
\resizebox{0.49\textwidth}{!}
{
$f=3\lambda \left (m-\frac{b+m}{a_{s}} \right ) \pm \sqrt{\frac{3a_{s}(2+3\lambda^{2})(b+m-ma_{s})^{2}-2\rho_{0,m}}{a^{3}_{s}}}$
}
\ee

\be 
d=\frac{1}{2}\lambda a_{s}h,
\ee

\be
\resizebox{0.49\textwidth}{!}
{
$c=\frac{\rho_{0,m}}{4a^{2}_{s}}-\frac{1}{2}(a_{s}-1)m(m-1)-\frac{[(a_{s}-1)m-b]^{2}}{a_{s}}-\frac{5}{4}\lambda f[(a_{s}-1)m-b].$
}
\ee

\be
h=-\frac{Af^{n-1}}{(n+1)\big(1+\frac{3}{2}\lambda^2 \big)}.
\ee

Notice that for $\rho_{0,m}=0$, all coefficients reduce to the ones in the absence of the fluid. Comparing them with the coefficients of quintessence models, we see that for $\lambda=0$ they reduce to them except for the coefficient $d$. This occurs because $d$ is the coefficient of the scale factor's diverging term. In quintessence models we have divergence of the third derivative of the scale factor, while in scalar-tensor models the second derivative of the scale factor diverges.

\subsection*{Proof of eq. (\ref{dotH})}

The scale factor and its first and second derivative are

\be 
a(t)=a_{s}+b(t_{s}-t)+c(t_{s}-t)^{2}+d(t_{s}-t)^{q},
\ee

\be \label{der1}
\dot a=-b-2c(t_{s}-t)-dq(t_{s}-t)^{q-1},
\ee

\be \label{der2}
\ddot a=2c+dq(q-1)(t_{s}-t)^{q-2}.
\ee

Close to the singularity eqs (\ref{scalefactor1}), (\ref{der1}), (\ref{der2}) become

\be \label{cs1}
a(t)=a_{s},
\ee

\be \label{cs2}
\dot a=-b,
\ee

\noindent and

\be \label{cs3}
\ddot a=2c
\ee

\noindent respectively.

\noindent Substituting eqs (\ref{cs1}), (\ref{cs2}), (\ref{cs3}) into the Hubble parameter and its derivative we have

\be \label{h1}
H=-\frac{b}{a_{s}}
\ee

\noindent and

\be \label{h2}
\dot H=\frac{2c}{a_{s}}-\frac{b^2}{a^{2}_{s}}
\ee

\noindent Substituting eqs (\ref{h1}), (\ref{h2}) in eq. (\ref{c-b}) we obtain

\be
\dot H=-3H^{2}.
\ee

\subsection*{Proof of eq. (\ref{dotHmatterred})}

Following the steps of the previous proof, in the absence of the fluid,
we have

\be \label{h1matter}
H=\frac{(a_{s}-1)m-b}{a_{s}}
\ee

\noindent and

\be \label{h2matter}
\dot H=\frac{(a_{s}-1)m(m-1)+2c}{a_{s}}-\frac{\left[(a_{s}-1)m-b\right]^{2}}{a^{2}_{s}}
\ee

Substituting eqs (\ref{cmatter}), (\ref{h1matter}) in eq. (\ref{h2matter}) we find

\be
\dot H=\frac{\rho_{0,m}}{2a^{3}_{s}}-3H^{2}.
\ee

and as a function of the redshift close to singularity $z_{s}$

\be 
\dot H=\frac{3}{2}\Omega_{0,m} (1+z_{s})^{3}-3H^{2}
\ee

\raggedleft

\bibliography{bibliography}

\end{document}